# *Message Passing Fluids: molecules as processes in parallel computational fluids*

Gianluca Argentini
gianluca.argentini@riellogroup.com

*New Technologies & Models*
*Information & Communication Technology Department*
*Riello Group, Legnago (Verona), Italy*

**Abstract**

*In this paper we present the concept of MPF, Message Passing Fluid, an abstract fluid where the molecules move by mean of the informations that they exchange each other, on the basis of rules and methods of a generalized Cellular Automaton. The model is intended for its simulation by mean of message passing libraries on the field of parallel computing. We present a critical analysis of the necessary computational effort in a possible implementation of such an object.*

**Key words**: cellular automaton, computational fluid dynamics, message passing, parallel computation.

## 1. Introduction

A *MPF*, Message Passing Fluid, is an abstract computational model of fluid into which the molecules move on the basis of some predefined rules which prescribe a mutual exchange of informations among the particles. A single molecule, once it has obtained these informations from the other neighbouring molecules, computes the value of the physical quantities prescribed by the rules of the MPF, and moves into the fluid on the basis of an appropriate combination of such values.

In a real fluid there are many physical quantities which determine the motion of the molecules. For the description of the intermolecular relations are yet determinant some short-range forces, above all those of electromagnetic nature, that the molecules exercise in a reciprocal manner by mean of an

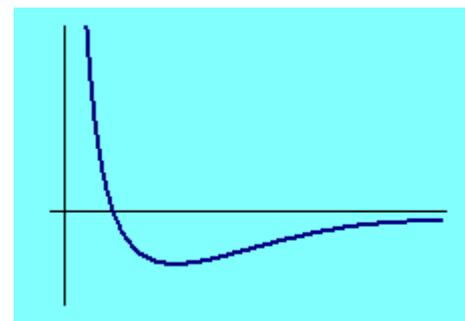

*exchange of informations*. For example, a force of electric nature very important in some fluid materials is the strength depending from the Lennard-Jones potential or from the Van der Waals potential, which determines an attractive action between particles sufficiently remote and a repulsive action when the distance is smaller than a definite value (v. p.e. [1], where is offered a description for computational purposes).

Therefore a MPF, which is based by definition on the exchange of messages among molecules, can be a tool for a computational interpretation of the intermolecular relations and for their influence on the total motion of the fluid.

MPF can be simulated by mean of a 2D or 3D computational grid, and the exchange of informations can be implemented by mean of opportune libraries as *MPI*, Message Passing Interface (v. p.e. [3]). In this manner the simulation can profit by the benefits of a parallel computing environment: as shown forward, the molecules are comparable to independent processes which in the same computational step exchange informations each other. Hence a MPF can be considered as a computational parallel environment.

This paper is a contribution to the study and testing of possible computational models and to the research for a description, by mean of Cellular Automata techniques, of the numerous real phenomena characterized by complexity in the matter of fluids mechanics. For a justification and a survey about the use of such methods on Computational Fluid Dynamics, one can see p.e. [2].

## 2. The abstract model

In the MPF model the fluid is described by a *n* x *m* grid of quadratic equal cells in the 2D case, and from a *n* x *m* x *h* grid of cubic cells in the 3D case. The cells where the molecules can move are inside a space delimited by walls, which are represented by geometric cells of the same kind too and that constitute the physical boundary into which the fluid is contained. The molecules can receive and send messages, which in the physical reality correspond to intermolecular forces fields. Even the single walls cells can send to molecules some messages, corresponding to forces fields especially of electromagnetic nature. In these first studied models we suppose that the walls don't receive messages from molecules. Therefore a MPF is first of all an object constituted by a grid, by a set of molecules, by a messages field among molecules and by another messages field, of different nature respect to the former, due to the possible presence of walls or obstacles. The messages that the molecules can exchange each other concern for example with the positions occupied on the grid, the potential values of the force field exercised in the interior of the fluid, the speeds and directions of motion. The walls or other objects can send messages about the values of the potential relating to the physical force that they exercise on the molecules. This force can be for example a repulsion of electric nature that obstructs the physical penetrability among molecules of not affine materials. For the messages exchange is useful to define a neighbourhood relative to a specific fluid molecule, made by the set of those molecules of the fluid itself and if necessary of those of the extra-fluid objects which can exchange messages with the involved molecule. For example, in the 2D case, such neighbourhood could be the circle centred on the molecule itself and having as radius a predefined value, as shown in Fig.1 .

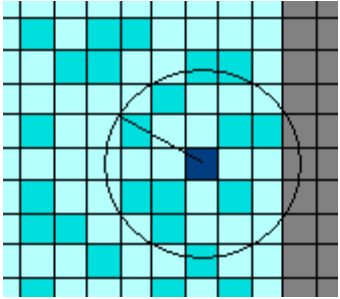

**Fig.1** *A possible circular neighbourhood of a molecule for determining the objects to which send messages at every computational step. With gray shade are represented the molecules of a wall or of an obstacle.*

Moreover one should define the rules by which one molecule moves itself, on the basis of the received messages. These rules are used for determining the cell of the grid where the molecule will move, after the receiving of the relative necessary informations by mean of the messages of the objects contained in its neighbourhood. For example, a possible rule in the 2D case could require the motion, among the eight cells adjacent to that where the molecule lies, to the cell where the sum of the potential values received via message is minimum. As supplementary rule one might require that, in the case that cell is already occupied, the molecule moves towards a cell immediately close to that determined before, on the basis of a random choice between right and left, or up and down.

This abstract MPF model leaves therefore a lot of free will for the choice of the fundamental characteristics: type of physical potentials, molecular neighbourhoods, motion rules. For every choice of these characteristics one obtains a particular MPF, which might have no correspondence with real physical situations; on the contrary for the possible study of a real fluid one should find the more convenient characteristics for implementing an adequate MPF model.

### 3. The computational model

One MPF has a native computational implementation by mean of techniques of messages passing among independent processes in a multithreading program. A first possible ideal computational accomplishment might be so modelled:

- every molecule corresponds to a process of the simulative program;
- the messages passing among the molecules is obtained by mean of interprocess calls to suitable functions; therefore the fluid is represented by a set of parallel processes;

- for every computational step, a molecule, hence a process, sends its data to the other molecules of the fluid, receives the data from the other objects contained in its neighbourhood, and then calcules and determines its shifting on the grid on the basis of the predefined rules;
- the grid is completly computed with the new positions of the molecules, and the program move to the next computational step.

Therefore a concrete computational realization of a MPF fluid requires the use of a message passing library in a parallel architecture program. The concrete cases that we have studied have been implemented by mean of the MPI library, interfaced to a `C` language program.

One might put forward the following remark: in a simulation with a high number of involved molecules, the number of processes should be the same as such amount, with consequent remarkable request of processors to obtain the desired degree of parallelism. This critical situation can be managed using a balanced distribution of molecules to every single process, hence to every single processor, for computing the new positions at every step.

A second remark might concern with the considerable number of messages managed by the processes. For example we assume that the fluid is composed by `N` molecules and that the number of processes is `P`, a divisor of `N`. At every computational step a process sends `P-1` messages (in MPI if we use the *broadcast* method the message is really only one), and it receives other `P-1`. Therefore at every step, as the number of processes is `P`, the total number of exchanged messages depends on $P^3$.

A third remark concern with the number of necessary calculations for determining the motion of the molecules towards the cells on the grid during a computational step. Fixed a molecule, for each step the process that managed it must check which from the `(N/P)-1` molecules of the process itself and which from the `(N/P)(P -1)` of the other processes belong to the predefined neighbourhood of the molecule; therefore, for every cell where the molecule itself can move, the process must compute the total potential due to the presence of the molecules of the neighbourhood, and finally find among those calculated the necessary value, on the basis of the predefined rules, for determining the motion. If we suppose that the neighbourhood of a molecule contains an average value `A`, constant for a fixed grid, of molecules, and that the number of the candidate cells for the new position of the molecule is `B`, we obtain that at every computational step the number of calculations for determining all the new positions on the grid is

$$N\left[1 + AB + \frac{N(P-1)}{P} + \frac{N}{P} - 1\right]$$

Therefore the number of calculation at every step depends on $N^2$.

In total the computational effort, for one step, is proportional to $N^2+P^3$. We now assume that the number of used processors is `P`, hence is equal to the number of processes launched at runtime by the simulation, and that the molecules of the flow at every step are `N=mP`. Then, as the processors can work in parallel, the time, in calculation unit, necessary for the execution of a complete computational step for the grid is proportional to

$$\frac{P^3 + N^2}{P} = P\ (m^2 + P)$$

and hence to $m^2$. Therefore to simulate a MPF the described algorithm, native and not optimized, presents a polynomial complexity of the second order.

## 4. An example

I present now a first simple example, from which one deduces that the MPF model can be an adequate tool for the description of the behaviour of a real fluid during the motion.

This model shows the trajectory of a single molecule in a fluid that move along a canal on a plane grid. Throughout its route there is a static obstacle. The motion of the molecule is regulated by the following rules:

i. Every neighbourhood centred on the molecule is circular and has a radius of 4 grid units;
ii. At every computational step the molecule can move only to one of the three upper cells immediately adjacent;
iii. The potential of an object that is in the position `(i,j)` of the grid, as regards the position `(p,q)` occupied by the molecule, is determined by mean of the following formula:

$$\frac{1}{|i-p| + |j-q|}$$

iv. the new position of the molecule is computed on the basis of the minimum of the total potential values due to the presence of the objects (walls of the canal, obstacle) internal to the neighbourhood centred on the molecule;
v. if the destination cell is already occupied, the molecule move to the cell having a potential value immediately bigger of the computed minimum; if the three upper cells are all occupied, the molecule doesn't move.

At every computational step the objects of the neighbourhood send a message to the molecule with an information relative to their position on the grid. The molecule computes for the upper free cells the total potential, given by the sum of the partial. Hence it moves on the cell where the potential is minimum. The trajectory of the molecule, which initially is placed on a cell adjacent to the wall, is shown by the following Fig.2:

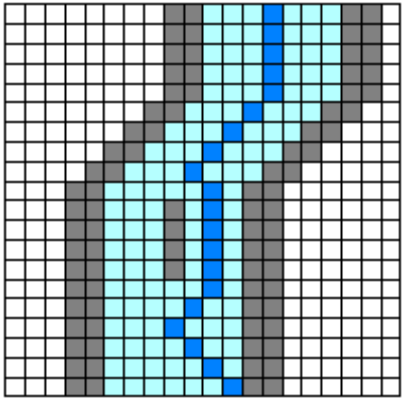

**Fig. 2** *Trajectory of a single molecule in the MPF material.*

When the molecule is close to the obstacle, it seems pushed far away, in agreement to the electrostatic repulsion forces. Moreover the trajectory respects the geometric shape of the canal into which the molecule moves. In this model the behaviour of the molecule seems to be sufficiently realistic to think that the MPF fluids can represent a promising research tool for the study of real fluid dynamics situations.

The described example has been implemented by mean of a `C` program interfaced to MPI calls. The program starts `P` independent processes; one of the processes traces in memory the position of the molecule on the grid and sends it to the other `P-1` processes by mean of the function `MPI_Bcast`; these processes are associated to the static objects and at every computational step transmit via `MPI_Send` to the process-molecule the positions of the object contained into the neighbourhood centred on the particle, whose process computes the relative potentials. The graphics is then obtained by the software `Mathematica`, which elaborates the two-dimensional array representing the grid, output of the `C` program. The Fig.2 is the result of a computation of about 10 seconds, with `P` = 4, on the Beowulf-like Linux cluster at CINECA, Bologna, Italy.

## 5. Possible further works

We are studying some implementations of the MPF model with a realistic number of molecules on an arbitrary geometry. The computation is made on the Linux cluster of Cineca. Moreover we are developing a possible generic library of MPI-like functions for the management of the messages in a MPF.

## 6. Bibliography

[1] *High Performances Computing II – Spring 2002*, Lectures, University at Buffalo, Department of Physics, web site: www.physics.buffalo.edu/phy516/lectures.html

[2] Norman MARGOLUS, *Emulating Physics: Cellular Automata that exhibit finite-state, locality, invertibility and conservation laws*, Computing Beyond Silicon Summer School, Caltech, 2002, web site: www.cs.caltech.edu/cbsss/schedule/slides/margolus1_emulating_physics.pdf

[3] Peter PACHECO, *Parallel programming with MPI*, Morgan Kaufmann, San Francisco, 1997

Gianluca Argentini, April 2003.

gianluca.argentini@riellogroup.com